\documentclass[apl, preprint]{revtex4}

\usepackage{graphicx}

\begin{document}

\title{Vector magnetic field microscopy using nitrogen vacancy centers in diamond}

\author{B.\,J.\,Maertz$^1$$^,$$^*$}
\author{A.\,P.\,Wijnheijmer$^1$$^,$$^2$$^,$$^*$}
\author{G.\,D.\,Fuchs$^1$}
\author{M.\,E.\,Nowakowski$^1$}
\author{D.\,D.\,Awschalom$^1$$^,$}
\email{awsch@physics.ucsb.edu}

\affiliation{$^1$\,Center for Spintronics and Quantum Computation, University of California, Santa Barbara, California 93106, USA
 \\
$^2$\,COBRA Inter-University Research Institute, Department of Applied Physics, Eindhoven University of Technology, P.O. Box 513, NL-5600 MB Eindhoven, The Netherlands
\\
$^*$\,contributed equally to this work}

\date{\today}

\begin{abstract}
The localized spin triplet ground state of a nitrogen vacancy (NV) center in diamond can be used in atomic-scale detection of local magnetic fields. Here we present a technique using these defects in diamond to image fields around magnetic structures. We extract the local magnetic field vector by probing resonant transitions of the four fixed tetrahedral NV orientations. In combination with confocal microscopy techniques, we construct a 2-dimensional image of the local magnetic field vectors. Measurements are done in external fields less than 50 G and under ambient conditions.
\end{abstract}

\maketitle

Visualizing magnetic field vectors has been an area of interest for many years. The familiar method of spreading iron filings around a bar magnet was a well practiced technique when Faraday first described ``lines of magnetic forces'' in 1831 \cite{Faraday}. In the past few decades, many techniques have been developed to locally probe much weaker magnetic fields on much smaller length scales.  Superconducting quantum interference devices\cite{Bending}, Scanning Hall probe microscopy \cite{Chang}, and magnetic resonance force microscopy \cite{Rugar, Mamin} are a few promising methods. Using the localized spin triplet ground state in a nitrogen vacancy (NV) center in diamond as an atomic-scale magnetic field probe has also been proposed \cite{Degen, Taylor}. Theoretical spatial and magnetic field resolution limits exceed those of previously mentioned techniques and inherent operation in ambient conditions and low external fields is also advantageous. Early experiments using NV centers as proximity magnetometers show promising results \cite{Balasubramanian, Maze, Acosta}. In this letter, we present a field imaging technique using NV centers in much the same way that Faraday used iron filings to view magnetic field lines.

\begin{figure}
\includegraphics{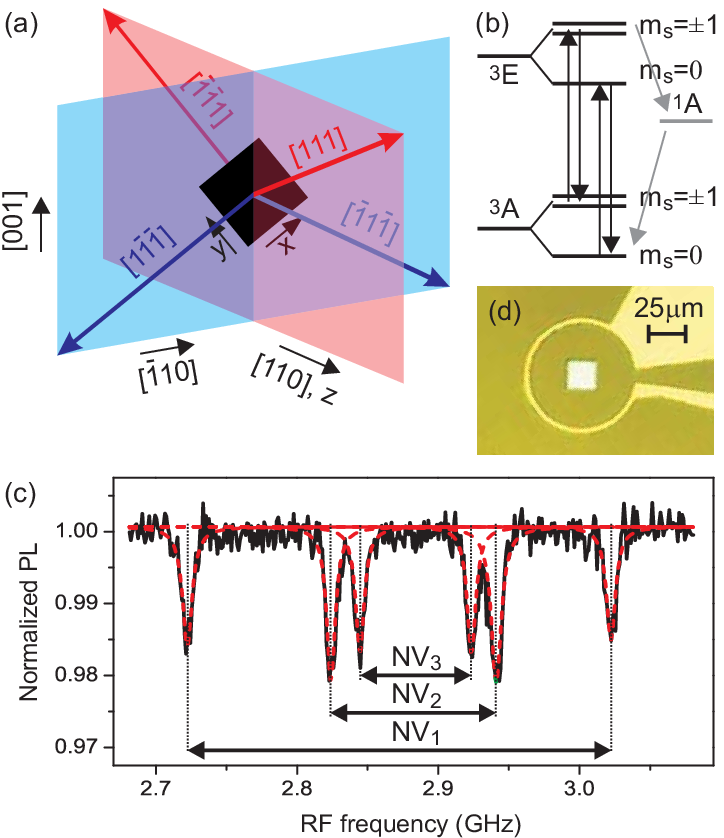}
\caption{\label{Fig1}
(a) Schematic of sample orientation showing the (110) diamond surface (blue plane) along with two NV symmetry axes (blue arrows). The red plane, perpendicular to the surface, contains the other two NV axes. The black square represents the patterned permalloy structure. Crystallographic directions of the diamond are depicted as well as the cartesian coordinate system. (b) Level structure of NV complex. (c) Typical ESR spectrum (solid black) and Lorenzian fits (dashed red). The splittings of the three NV directions are extracted from fits. (d) Optical image of the sample showing $20 \times 20$ $\mu$m permalloy square with 75\,$\mu$m diameter microwave antenna on diamond.}
\end{figure}

A NV center in diamond consists of a substitutional nitrogen adjacent to a vacancy in the lattice. The symmetry axis is along any of four the tetrahedral $\langle 111 \rangle$ crystallographic directions, see \mbox{Fig.\ \ref{Fig1}(a)}. Two of the symmetry axes lie on the $(1 1 0)$ plane (blue) and the other two on the $(\bar{1} 1 0)$  plane (red). The pertinent level structure associated with the negatively charged NV center used in this experiment is depicted in \mbox{Fig.\ \ref{Fig1}(b)}. The NV forms a spin triplet in the ground state ({\it $^3$A}) and excited state ({\it $^3$E}). The degenerate \mbox{m$_s = \pm 1$} states are zero-field split from the m$_s=0$ state by 2.87\,GHz.  Optically addressing the system induces spin conservative transitions from {\it $^3$A} to {\it $^3$E}. However, a non-radiative relaxation mechanism ({\it $^1$A}) allows optical initialization into the \mbox{$m_s$\,$=$\,$0$} state \cite{Jelezko,MansonPRB}. This shelving state causes a decreased photoluminescence (PL) intensity when the system is in the \mbox{$m_s$\,$=$\,$\pm1$} states. The degeneracy between the m$_s = \pm 1$ states is lifted by applying a magnetic field, which Zeeman splits the \mbox{m$_s = \pm 1$} states and does not affect the \mbox{m$_s = 0$} state. Applying a microwave field drives transitions between the \mbox{m$_s = 0$} state and the \mbox{m$_s = +1$} \mbox{(-1)} state when the frequency is resonant with the splitting between the \mbox{m$_s = 0$} state and the \mbox{m$_s = +1$} \mbox{(-1)} state. This results in Lorentzian dips in the optically detected electron spin resonance (ESR) spectrum at the resonant frequencies (see \mbox{Fig.\ \ref{Fig1}(c)}) \cite{Loubser}. Three distinct Zeeman split pairs of peaks originate from three different NV orientations($[1 \bar{1} \bar{1}$], $[1 1 1$], and $[\bar{1} \bar{1} 1$] in \mbox{Fig.\ \ref{Fig1}(a)}); the NV along the fourth orientation ($[\bar{1} 1 \bar{1}$] in \mbox{Fig.\ \ref{Fig1}(a)}) is difficult to resolve due to optical polarization and gives degenerate information.

We use (110) diamond substrates (Sumitomo type Ib grown by high-temperature high pressure methods) with sufficiently high concentrations of NV's to have all four orientations within the confocal volume \mbox{($\sim1$\,$\mu$m$^3$)} of our microscope. Circular microwave antennas are patterned on the diamond, defining the magnetic sensing region. For this study, we use antennas that are 75 $\mu$m in diameter. Permalloy (Ni$_{0.8}$Fe$_{0.2}$) shapes, lithographically aligned with respect to one in-plane NV, are thermally evaporated in the center of the antennas with a thickness of \mbox{$\sim50$\,nm}, see \mbox{Fig.\ \ref{Fig1}(d)}. Three different shapes were investigated in this work: a \mbox{$20 \times 20$ $\mu$m} square, an equilateral triangle with \mbox{20 $\mu$m} sides, and a \mbox{$10 \times 40$ $\mu$m} rectangle.

In order to extract the local magnetic field vector, we measure the Zeeman splitting for the three visible NV's. We apply an external field with a permanent magnet on a rotation and translation stage. The range of motion leads to fields between \mbox{$45$\,G} and \mbox{$1000$\,G}. The field is reversed by inverting the magnet. A 532\,nm laser is focussed to a spot size of \mbox{$0.3$\,$\mu$m$^2$} a few microns below the diamond surface. This initializes the spin state of the NV's within the confocal volume to the m$_s = 0$ state. A fast steering mirror is used to spatially scan across the sample with a 2-dimensional scanning range of \mbox{$20 \times 20$\,$\mu$m$^2$}. Larger movements are done using translation stages. The PL intensity is measured with an avalanche photodiode (APD). A signal generator is used in combination with on-chip waveguides to apply microwave fields for ESR measurements.

Each ESR spectrum is fit, and the frequency splittings are extracted. The splitting is proportional to the projection of the magnetic field along that NV's symmetry axis \cite{note}. Using simple geometric arguments to transform the tetrahedral directions into cartesian coordinates (as defined in \mbox{Fig.\ \ref{Fig1}(a))}, we find the following equations for the magnetic field components

\begin{figure}
    	\includegraphics{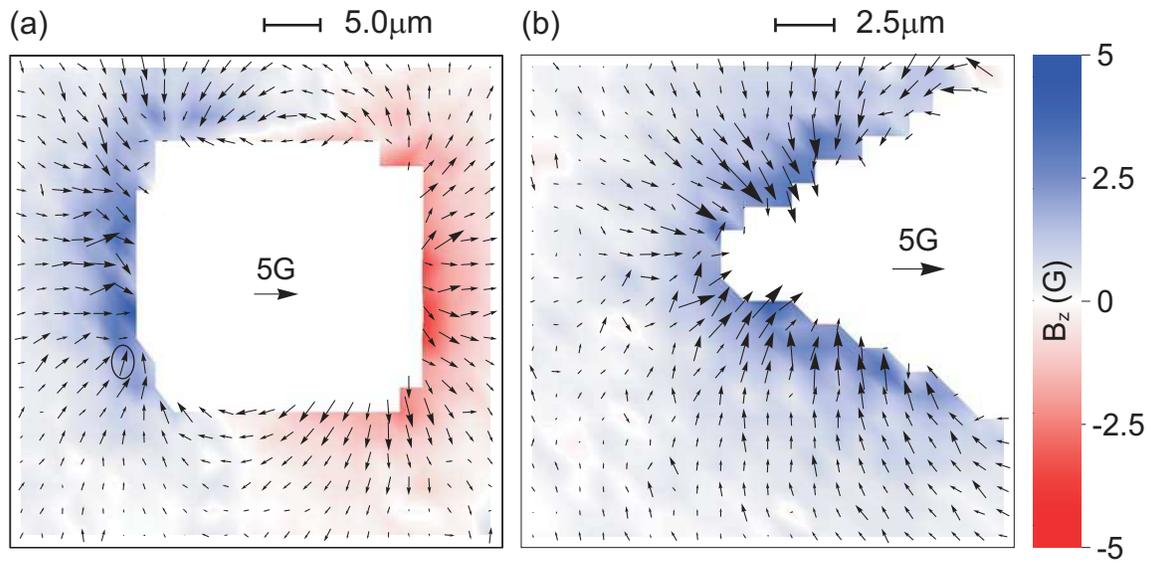}
      \caption{Images of the magnetic field vectors around the permalloy (a) square and (b) triangle measured in an external field of \mbox{45\,G}. Arrow's size and direction represent the x-y vector, while z is depicted with color. Each vector refers to one ESR spectrum. To get a sense of scale, the circled field vector is $[1.36 \pm 0.13, 2.48 \pm 0.22 , 1.76 \pm 0.09]$\,G.
      \label{Fig2}}
\end{figure}

\begin{tabular}{l}
$B_x = \beta \cdot \Delta \mbox{NV}_1 $\\
$B_y = \beta \cdot \frac{1}{2 \sqrt{2}} \left( 2 \ \Delta \mbox{NV}_1 -3 \left( \Delta \mbox{NV}_2 + \Delta \mbox{NV}_3 \right) \right) $\\
$B_z = \beta \cdot \frac{\sqrt{3}}{2 \sqrt{2}} \left( \Delta \mbox{NV}_3 - \Delta \mbox{NV}_2\right)$,
\end{tabular}

where $\beta = \frac{h}{2 g \mu_B}$, $\Delta $NV$_i$ is the splitting between the m$_s = +1$ and the m$_s = -1$ state of NV$_i$; $i=1$ corresponds to the visible in-plane NV and $i=2,3$ correspond to the two out-of-plane NV's. By measuring ESR spectra while scanning the laser spot over the sample, we determine the splittings of NV$_i$. We are then able to extract the magnetic field vector at each point. A reference ESR spectrum is measured far away from the permalloy, allowing us to subtract the externally applied field from the total field in order to get the local field. The resulting images of the local field are shown in \mbox{Fig.\ \ref{Fig2}}.

In our measurement, we cannot distinguish between the m$_s = +1$ and the m$_s = -1$ state, or in other words, between positive and negative splittings. This results in an ambiguity in the sign of the magnetic field. However, by using a subtractive method, we measure small changes with respect to the uniform external field. With careful choice of the external field, the ESR dips do not cross as we scan across the sample. Since we know the external field, we can unambiguously extract the sign of the local field.

\begin{figure}
\includegraphics{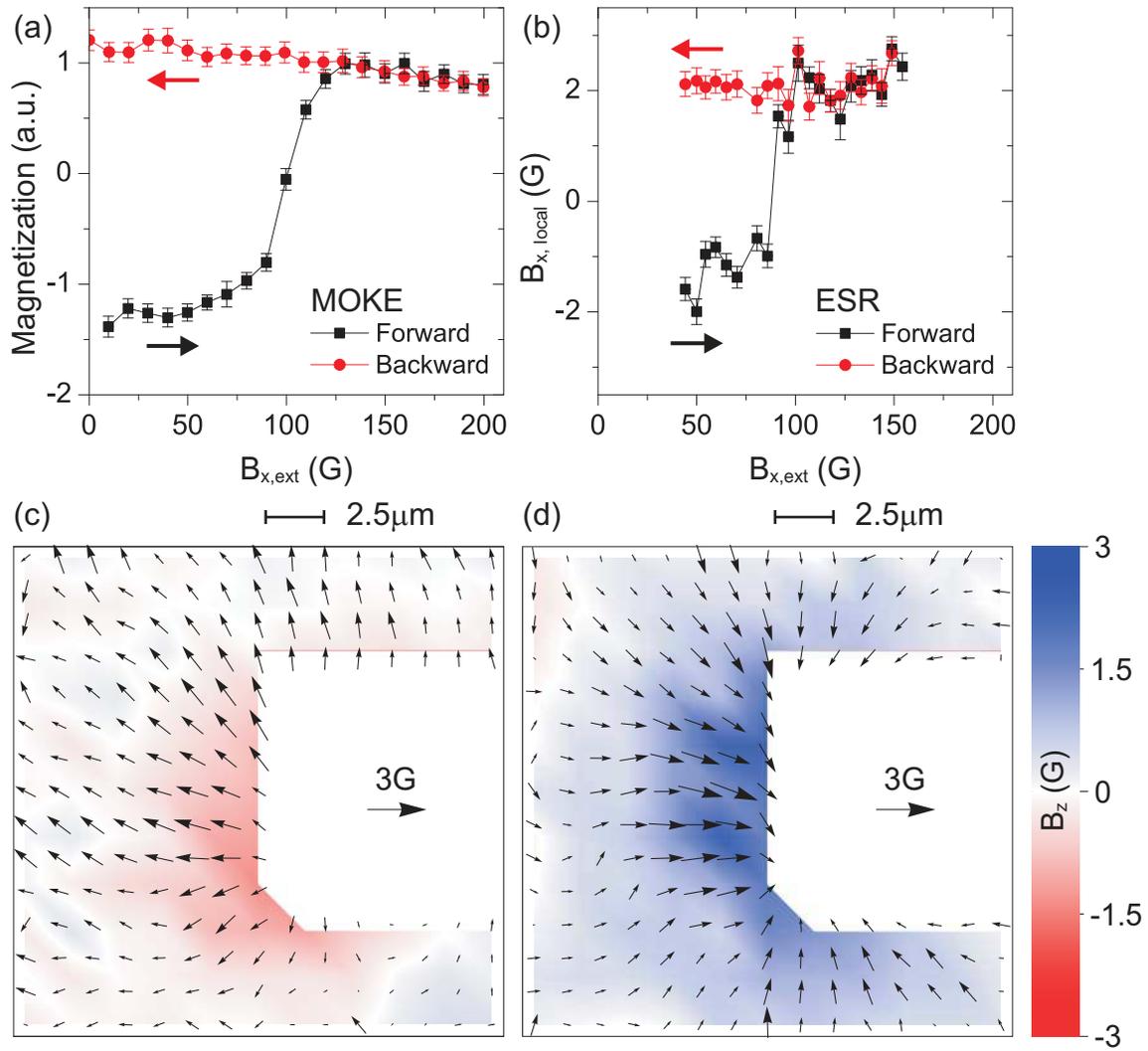}
\caption{\label{Fig3}
Hysteresis of the permalloy rectangle as measured by MOKE (a) and ESR (b). Increasing (decreasing) magnetic field sweep is in black (red).
Vector field image of the field lines around the rectangle with the
magnetization in the negative (c) and the positive x-direction (d).}
\end{figure}

The magnetization of the permalloy is characterized in a Magneto-Optical Kerr Effect (MOKE) microscope. The hysteresis of the magnetization along the length of the rectangle is shown in \mbox{Fig.\ \ref{Fig3}(a)}. The MOKE measurement reveals a coercive field of \mbox{$\sim$\,100}\,G along the long axis.

Magnetization of the permalloy is also investigated using ESR to probe the field around the rectangle. We move to a spot next to the side of the rectangle and prepare the permalloy in an anti-parallel configuration by first applying a large negative field (\mbox{-\,1000\,G)} and subsequently reducing and inverting the field to \mbox{+\,45\,G}. While measuring ESR spectra, we step the magnetic field to \mbox{+\,155\,G} and back to \mbox{+\,45\,G}. We also measure ESR spectra far away from the permalloy at each field step as a reference. By subtracting the reference magnetic field from the total magnetic field, we find the local magnetic field originating from the permalloy. The resulting data (\mbox{Fig.\ \ref{Fig3}(b)}) reproduces the hysteretic behavior seen in the MOKE data. It is important to note that this measurement does not directly show the magnetization, but shows the x-component of the local magnetic field due to the magnetization of the permalloy.

Images of the magnetic field vectors before the forward sweep and after the backward sweep are shown in \mbox{Fig.\ \ref{Fig3}(c) and (d)} respectively. Both images were taken in the same external field of \mbox{50\,G}. The reversal in direction of magnetic field vectors, and thus in the magnetization, is clearly seen.

The magnetic field and lateral resolutions of the measurements presented here are $\sim0.2$\,G and $\sim0.3$\,$\mu$m respectively, which can be improved by implementing a few techniques. Improvement of the magnetic field resolution up to \mbox{$0.3$\,mG} can be done by using pulsed techniques \cite{Taylor,Maze}. The spatial resolution can be significantly improved by implementing a reversible saturable optically linear fluorescence transitions (RESOLFT) technique \cite{Hell}, such as stimulated emission depletion (STED) microscopy, which has been shown to improve imaging resolution of diamond NV's to a few nm \cite{Rittweger}.

In summary we showed that NV centers in diamond can be used to spatially image local magnetic field vectors. The resolution in this work is $\sim0.2$\,G and $\sim0.3$\,$\mu$m. The important advantages of the technique presented here are simultaneous extraction of three orthogonal components of the magnetic field and operation in small external fields less than $\sim50$\,G and under ambient conditions.

We thank F.\ J.\ Heremans for measurement preparation and AFOSR and ARO for their financial support. A portion of this work was done in the UCSB nanofabrication facility, part of the NSF funded NNIN network.

\newpage

\begin{thebibliography}{15}
\bibitem[{\citenamefont{note}(2009)}]{note}
	\bibinfo{author}{\bibfnamefont{This does not hold when \mbox{$\overrightarrow{NV} \perp \overrightarrow{B}$} where the \mbox{$m_s$\,$=$\,$\pm1$} are still split, even though the projection of $\overrightarrow{B}$ is 0. In this work, this situation is avoided with careful choice of external fields.}}
    \bibinfo{journal}{} \bibinfo{volume}{}
    \bibinfo{pages}{} \bibinfo{year}{}
\end{thebibliography}
\end{document}